# Term-Mouse-Fixations as an Additional Indicator for Topical User Interests in Domain-Specific Search

Daniel Hienert and Dagmar Kern
GESIS – Leibniz Institute for the Social Sciences
Cologne, Germany
firstname.lastname@gesis.org

## ABSTRACT

Models in Interactive Information Retrieval (IIR) are grounded very much on the user's task in order to give system support based on different task types and topics. However, the automatic recognition of user interests from log data in search systems is not trivial. Search queries entered by users a surely one such source. However, queries may be short, or users are only browsing. In this paper, we propose a method of term-mouse-fixations which takes the fixations on terms users are hovering over with the mouse into consideration to estimate topical user interests. We analyzed 22,259 search sessions of a domain-specific digital library over a period of about four months. We compared these mouse fixations to user-entered search terms and to titles and keywords from documents the user showed an interest in. These terms were found in 87.12% of all analyzed sessions; in this subset of sessions, per session on average 11.46 term-mouse-fixations from queries and viewed documents were found. These terms were fixated significantly longer with about 7 seconds than other terms with about 4.4 seconds. This means, term-mouse-fixations provide indicators for topical user interests and it is possible to extract them based on fixation time.

## 1 INTRODUCTION & RELATED WORK

Knowing the user's search task and the current interest would be very valuable for supporting the actual information need. However, this is still a challenging issue in real-world and live situations. There are various research attempts to model and predict user needs and interests, for example, based on user queries [e.g. 7], context [e.g. 11] and search histories [e.g. 9, 10]. As eye-tracking data is still not practical in long-term real-life user studies, we logged the position of the mouse on a term as well as its dwell time and used this as indicators for user's interest in this term.

Mouse-movement has been shown to be a promising candidate for gathering further information about user behavior. Mouse trajectories, for example, are utilized to infer and disambiguate navigation and informational search intents [3]. Huang et al. [6] found that cursor hovering and scrolling on landing pages are good indicators to decide if a user has examined a search result. Mouse movement and scrolling are used in addition to click-through-rate and dwell time to better estimate document relevance [2]. The approaches above focus thereby on areas of interests on result lists and landing pages. The approach by Ageev et al. [1] goes one step further and considered single fixated terms on the landing pages and used them among other things for generating result summaries of the corresponding document. Liu et al. [8] combine existing click models with mouse movement information to enhance the prediction of result examination. Therefore, they collected a large-scale data set with a commercial search engine. Other studies in this context base their findings predominantly on task-based evaluations in laboratory settings.

In our research, we also refer to real-world interaction data collected in a digital library for social science information and focus on mouse-fixated terms in whole user sessions. We address the following research questions:

*R1*: Can we find indicators of topical user interests such as user search terms and topics from document clicks in mouse-fixated terms?

*R2*: Is it possible to distinguish between terms in a list of mouse-fixated terms the user showed an interest in and terms the user had fixated more or less unconsciously?

With our work, we contribute to this research field by analyzing a log file of about 22,000 search sessions of a domain-specific digital library.

## 2 EXPERIMENT

### 2.1 Environment

Sowiport[1] is a digital library for social science information such as bibliographic records, full texts, and research projects. It contains more than nine million records from 22 German- and English-language databases; the main audience is German-speaking. Users are supported in their search process with a number of services [cp. 4]. Figure 1 shows the search result page of Sowiport. By clicking on the title of a result entry the user is forwarded to the corresponding detailed view page (see Figure 2). This page contains further information about the selected bibliographic record. Amongst typical literature metadata like

---

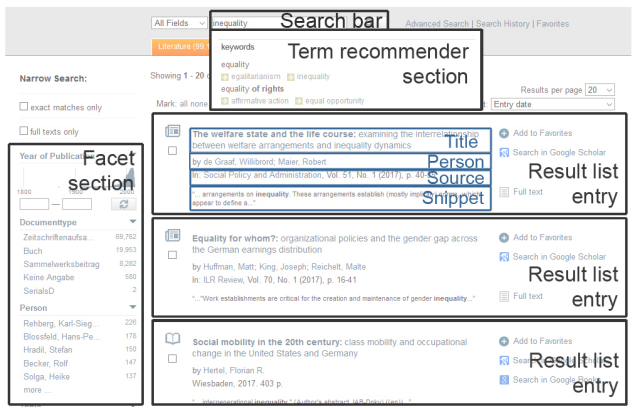

**Figure 1: Overview of the different AOIs on the search result page (black frame) with metadata fields (blue frame).**

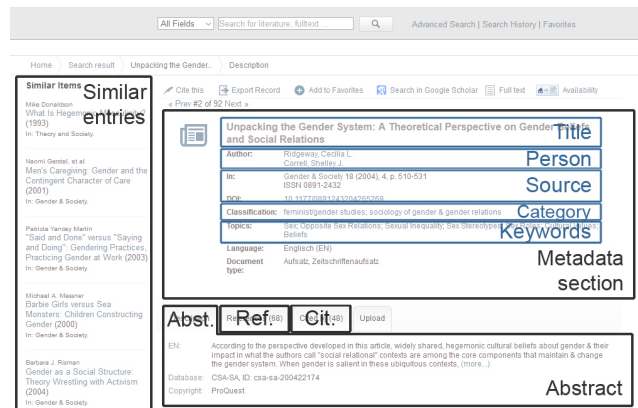

**Figure 2: AOIs on the detailed view page with metadata fields.**

author and source, it usually provides keywords and sometimes even categories. These keywords and categories are manually assigned from information professionals using the thesaurus for the social sciences[2] (TheSoz), the classification for the social sciences[3] and similar classification systems.

## 2.2 Mouse Tracker

We have implemented our own JavaScript mouse tracker and integrated it in Sowiport to capture aggregated mouse fixations over terms. The framework can be customized to capture only mouse fixations in certain areas of interest (AOI). In our case, we limited the recording of the search result page to the term recommender section, the result list entries (with the metadata fields: title, person, journal/proceedings source, and snippet), and the facet section. The recording of the detailed view page is limited to the metadata section (with the metadata fields: title, person, journal/proceedings source, category, and keywords), abstract, references, citations and similar entries section. Figure 1 gives an overview of different AOIs and metadata fields on the search result page and Figure 2 on the detailed view page.

When the user hovers with the mouse over a word in these areas the algorithm creates a new entry for this term in the term-mouse-fixation data file[4]. To achieve a good aggregation level already in this stage words are cleaned from English and German stop words and stemmed with a Porter stemmer. Fixation times for each stemmed term are summed up for every time the user hovers over an instance on the whole website. This means, the fixation time describes how long a user has fixated a unique term with the mouse over the whole session. For each term, the aggregated fixation time, the first-time-fixation, last-time-fixation, the AOI and metadata field are recorded. Each time the user submits a search query with the search form the term-mouse-fixation data is stored in the logging database together with the session-id, timestamp and user search terms.

## 2.3 Methodology for User Queries

In a first step, we check for the correspondence of term-mouse-fixations to *user search terms*. The overall goal is to check whether user search term(s) from user queries have been fixated with the mouse in the search session at all. For each user session and its user queries a list of distinct user search terms is built. User search terms with more than two characters are cleaned from English/German stop words and stemmed by a Porter stemmer. The list of term-mouse-fixations is additionally cleaned with a blacklist of terms which are part of the user interface and have no substantive topical meaning (e.g. "Authors:"). Then, the algorithm compares each user search term to the list of term-mouse-fixations. The comparison checks for in-word-inclusion, this means user search terms in the middle of term-mouse-fixations are also recognized. This is especially important for the German language where a lot of compounds are used. Found user search terms in term-mouse-fixations are collected throughout the user session. Based on this we compute the average fixation time of all found user search terms and compare it to the average fixation time of terms in term-mouse-fixations which are no user search terms. Additionally, we analyze the source AOIs and metadata fields for found search terms in term-mouse-fixations.

## 2.4 Methodology for Document Clicks

Similarly, we check for the correspondence of term-mouse-fixations to topics from *documents clicks*. We define a *document click* as a click in a result list entry that leads to further information about the document or to the document itself. These are mainly clicks on the title to see the detailed view (Figure 2) of a document within Sowiport and clicks on elements in the sidebar of a result item which lead to the full text outside Sowiport. We assume that these clicks indicate a certain user interest in this document. For each *document click* we collect the metadata of the corresponding document. As documents in our collection are well described by title and keywords we focus in the following on these fields. Titles are tokenized, stop word cleaned and like the keywords stemmed. The algorithm then compares if either document keywords or title terms can be found within the term-mouse-fixations. Found title terms and

---

keywords are collected for the whole user session. Again, for each session, we compute the average fixation times for found title terms and keywords and compare them to the average fixation time of those mouse-term-fixations which are no title terms or keywords.

## 3 RESULTS

### 3.1 Evaluation Data Set

The final data set has been recorded from 18[th] October 2016 to 13[th] February 2017. User sessions are limit to those with at least one submitted search query which results in 22,259 sessions with 80,796 searches and 105,286 document clicks. On average a session lasts 64 minutes, and about 57 distinct terms have been fixated within the session with the mouse. Figure 3 shows the distribution of first-time mouse-fixation by AOIs and metadata fields. More than half of the fixated terms are first hovered in the result list entries (58.46%), followed by the metadata section in the detailed view (21.11%), the facets (9.21%) and by other AOIs. The metadata fields title (25.79%), person (24.19%) and snippet (21.99%) are relatively evenly distributed.

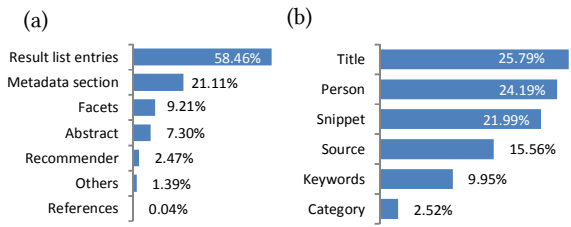

**Figure 3: Proportion of first-time mouse-fixation in (a) AOI and (b) metadata field.**

### 3.2 User search terms in term-mouse-fixations

Now we can check how many of the search terms the user has explicitly entered into the search bar can be found in term-mouse-fixations.

About half of the user search terms (47.41%) have been fixated with the mouse within the session. We call this proportion "*found terms*" in term-mouse-fixations. Inversely, "*other terms*" are the proportion in term-mouse-fixations which have no correspondence to user search terms. *Found terms* have been fixated significantly longer with the mouse (9.11 seconds) than *other terms* (4.41 seconds). These are statistically significant different groups found with a single factor ANOVA test with $\alpha=0.01$ and $F=1,736.13$.

Figure (4a) shows the AOI where the *found search terms* come from based on first fixation. Most terms have been fixated in the result list entries (49.45%), followed by the metadata section (23.68%), the term recommender section (15.46%), the facets (4.85%), and in the abstract (4.63%). Figure (4b) shows the distribution of metadata fields in which terms have been fixated. 26.93% came from the title, 8.84% from the snippet, 8.84% from the keywords, 6.16% from the persons, 5.20% from the source and 1.81% from the category. The rest to 100% has no metadata field information.

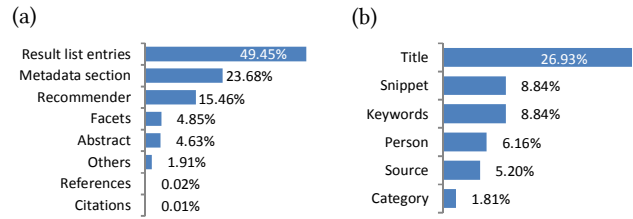

**Figure 4: Source of found search terms in term-mouse-fixations: (a) AOI and (b) metadata field.**

### 3.3 Document topics in term-mouse-fixations

Next, we check for correspondence of term-mouse-fixations with topics from *document clicks* in the result list.

First, we compare term-mouse-fixations with title terms from clicked documents and found that for 72.13% of the clicked documents at least one title term is found in the term-mouse-fixation of the appropriate session. Regarding the number of found title terms on average 2.48 of 4.05 terms (61.23%) of a single document are found. Fixation times are significantly longer with 8.79 seconds for *found title terms* vs. 4.32 seconds for the rest of term-mouse-fixations (significantly different with ANOVA, $\alpha=0.01$ and $F=1786.68$).

For the *keywords* of the clicked documents, we found a similar result. For 75.06% of all document clicks at least one keyword can be found in the term-mouse-fixations of the appropriate session. Regarding the number of found keywords on average 3.51 of 11.4 keywords (30.79%) of a single document can be found in term-mouse-fixations of the appropriate session. Found keywords have been fixated significantly longer with 6.42 seconds than other term-mouse-fixations with 4.33 seconds (significantly different with ANOVA, $\alpha=0.01$ and $F=563.84$).

### 3.4 Combined results

Finally, we can combine the check for inclusion of user search terms, title terms and keywords from document clicks in term-mouse-fixations within the same user session. We then find on average 11.45 terms in 87.12% of the sessions. Found terms have an average fixation time of 7.06 seconds, other terms 4.43 seconds (significantly different with ANOVA, $\alpha=0.01$ and $F=1,296.30$).

## 4 DISCUSSION

So far, mouse movements, clicks, and scrolling data have been used as an articulation of user behavior to understand the user's interest for e.g. certain areas or documents. In this paper, we go one step deeper and consider the term under the mouse cursor as a point of interest. In a prior eye tracking study, we found that for a specific exploratory task users are scanning the user interface with their eyes for new search terms and later use them in their search queries [5]. These terms have been fixated several times before they were used as search terms. As eye tracking is still not available in real world settings, we want to know if users' mouse pointer behavior can provide enough information to assume topical user interest. Therefore, we formulated the following research question: *(R1) Can we find indicators of topical user interests such as user search terms and topics from document clicks in mouse-fixated terms?*

User search terms are a very concrete and condensed articulation of user interests, and we found that in almost half of the user sessions (47.41%) the user search terms can be found in the term-mouse-fixations. If we compare Figure 3 and Figure 4 we can see that the metadata fields person, snippet, and source are often first-time hovered with the mouse, but the most user search terms originated from title and keywords. This is analog to our results in the eye tracking experiment [5] where user search terms have been most times fixated in these fields. Document clicks are a second indicator for analyzing user interest. Here, we focus on the two most preferred sources for user search terms – title and keywords. We found that for 72.13% of the documents, the user showed an interest in, at least one title term and in 75.06% of all clicked documents at least one keyword is found in term-mouse-fixations. On average 2.48 title terms and 3.51 keywords per single document are represented. We assume this as a reasonable representation of the document's topic itself and therefore for the user's topical interest. If we combine these findings, we can find in 87.12% of the sessions 11.45 terms from user queries and topics of clicked documents.

For a practical use of our previous findings, it is essential that we can determine mouse-fixated terms the user shows an interest in. On average 57 terms have been fixated in a search session, but only a share of them can be used to represent topical user interests. We address this issue in our second research question: *(R2) Is it possible to distinguish between terms in a list of mouse-fixated terms the user showed an interest in and terms the user had fixated more or less unconsciously?* We found a strong proof that accumulated fixation times can be used to extract these terms. Fixation times are significantly longer for found search terms (9.11s), keywords (6.42s) and title terms (8.79) from document clicks. The fixation times for the rest of term-mouse-fixations are very stable at around 4.4s. With this knowledge, we have one indicator to distinguish between important and unimportant terms in the sense of user interests. To make the prediction for user topical interest more precise, we consider using term overlaps between documents. The general assumption here is that keywords which occur simultaneously in several clicked document of a user's search session express even more strongly her interest. Based on the evaluation data set we computed the fixation times of keywords which occur in two to five different clicked documents of a search session. For keywords in x documents of a session, we found fixation times from 6-10s for keywords (2 docs: 6.69s, 3 docs: 7.79s, 4 docs: 8.84s, 5 docs: 9.74s). Again, the fixation times for other terms in term-mouse fixation remain stable at around 4.4 to 4.8 seconds. This means, even terms representing indicators of stronger user interest (contained in several different documents of a session) show linear higher fixation times. This is a further indication that term-mouse-fixation times can be used for the extraction of interesting terms.

Nevertheless, the approach of term-mouse-fixations also has some limitations. Mouse moving behavior can be very individual for a single user. For example, while one user extensively moves the mouse for reading assistance, the other user does not. This can result, e.g., in large deviations of fixation times and may influence the quality of extracted topical user interest on single session level.

## 5 CONCLUSION & FUTURE WORK

In this work we analyzed logfiles of a domain-specific digital library and we found that in 87% of the sessions we can find 11.45 terms per session from user queries, title, and keywords from clicked documents in term-mouse-fixations. These terms have significantly longer term-mouse-fixation times (7.06s) than the rest of the term-mouse fixations (around 4.43s). With the difference in fixation time, we can extract these terms from the whole list of term-mouse-fixations. These terms are indicators for user interests articulated through search queries and document clicks. One current line of research in IIR is to find the user's task(s) within a search session. The type of the task and the topic can help to better support the user in different search situations and for different search topics. Different sources of background knowledge can be used to understand these topics such as user queries, actions, the context or history. In our research, we found with term-mouse-fixations an additional source of information to understand the user's topics. Combining different sources may lead to a better estimation of topical user interests. In the end, we could rank search results according to the extracted user interest or more personalized recommendations could be given. In future work, we want to determine the quality of the assumed user interest based on keywords and title terms by performing a long-term field study with Sowiport in which we offer the extracted topics from term-mouse-fixations as suggested search terms.

**ACKNOWLEDGMENTS** This work was partly funded by the DFG, grant no. MA 3964/5-1; the AMUR project at GESIS.